\documentclass[conference]{IEEEtran}

\usepackage{cite}
\usepackage{amsmath,amssymb,amsfonts}
\usepackage{amsthm}
\usepackage{graphicx}
\usepackage{booktabs}
\usepackage{siunitx}
\usepackage{xcolor}
\usepackage{multirow}
\usepackage[colorlinks=true,allcolors=blue]{hyperref}
\usepackage{arydshln}

\newtheorem{problem}{Problem}

\newcommand{\context}{c}
\newcommand{\contextset}{\mathcal{C}}
 
\DeclareMathOperator*{\argmin}{arg\,min}

\title{Context-Conditioned Hamilton–Jacobi Reachability for Adaptive Safety Filtering}

\author{
    \IEEEauthorblockN{
        Ali Fuat Sahin\IEEEauthorrefmark{1},
        Yunus Yazoglu\IEEEauthorrefmark{3},
        John Talbot\IEEEauthorrefmark{2},
        Zeyuan Feng\IEEEauthorrefmark{3},
        James Dallas\IEEEauthorrefmark{2}, and
        Somil Bansal\IEEEauthorrefmark{3}
    }
    \IEEEauthorblockA{
        \IEEEauthorrefmark{1}École Polytechnique Fédérale de Lausanne, Email: alifuat.sahin@epfl.ch\\
        \IEEEauthorrefmark{2}Toyota Research Institute, Email: \{john.talbot, james.dallas\}@tri.global\\
        \IEEEauthorrefmark{3}Stanford University, Email: \{yazoglu, zeyuanf, somil\}@stanford.edu
    }
}

\begin{document}
\maketitle
\thispagestyle{empty}
\pagestyle{empty}

\begin{abstract}
Hamilton--Jacobi reachability constructs safety certificates for specified dynamics and safety constraints, tying each certificate to the deployment context for which it is synthesized. We ask whether a single certificate can instead represent a family of context-dependent safety problems and be queried across deployment conditions without re-synthesis. We learn a backward reachable tube for an eight-state vehicle model conditioned on local boundary geometry, friction coefficient, and adversarial disturbance scale. Geometry enters through an ego-frame boundary observation that defines the local containment constraint, while friction and disturbance scale enter as explicit operating-condition variables. This allows the same value function to be queried across friction coefficients from 0.4 to 2.0 and on geometries absent from synthesis. On 11 held-out evaluation geometries, the certificate maintains containment across the full tested friction range, including simultaneous geometry and grip shifts, while remaining within 1.2 percentage points in intervention rate and 0.09~m/s in speed of certificates re-synthesized with knowledge of the test geometry. We then deploy the certificate as a sampled discrete-time control barrier function filter on a full-scale vehicle near the handling limit. Lateral containment holds in every hardware session under both adversarial driving and autonomous racing, with 99th-percentile acceleration magnitude reaching 0.99~g. Across three certificates evaluated under a fixed autonomous racing controller, lap time varies by only 3.1\%, demonstrating that a context-conditioned reachability certificate can transfer to deployment geometries absent from synthesis with modest performance cost.
\end{abstract}

\section{Introduction}

Safety-critical control depends on knowing not only which states are safe, but under \emph{which conditions} they are safe. For a vehicle operating near the limit of tire adhesion, those conditions can change substantially across deployments: track geometry determines the admissible region, while tire--road friction determines which states can still be recovered within that region. A safety certificate synthesized for one deployment is therefore tied to the context in which it was constructed.

Hamilton--Jacobi (HJ) reachability makes this dependence explicit. Given a dynamics model, safety constraint, and disturbance model, HJ reachability computes a value function whose zero level set separates states from which the constraint can be maintained from those from which it cannot~\cite{mitchell2005,bansal2017hj}. Changing the deployment context can consequently change the reachable set itself. For a vehicle, this occurs not only when the track changes, but also when the available grip differs from the value assumed during synthesis. In particular, a certificate synthesized under optimistic friction can overestimate the set of recoverable states when deployed on lower-grip pavement. This raises a broader question: \emph{can a single value function represent a family of context-dependent reachability problems, rather than requiring a new synthesis for every deployment condition?} 

We learn a backward reachable tube for an eight-state vehicle model conditioned on local boundary geometry, friction coefficient $\mu$, and adversarial disturbance scale $s_d$. Geometry is provided as an ego-frame boundary observation, allowing the same value function to be queried at runtime on geometries absent from synthesis. This formulation makes it possible to distinguish the \emph{context assumed by the certificate} from the conditions encountered by the vehicle, providing a direct way to study transfer under geometry and grip mismatch.

\begin{figure}[t]
\centering
\includegraphics[width=\columnwidth]{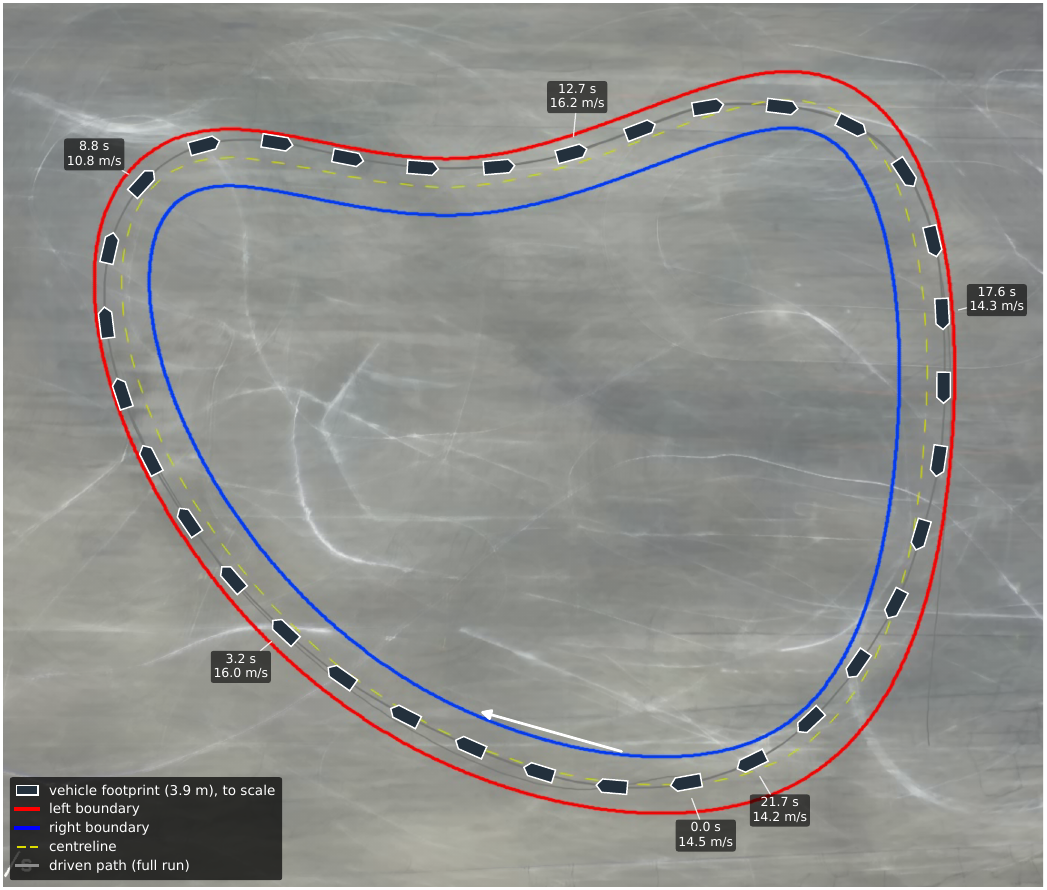}
\caption{Filtered closed-loop operation at the deployment site. The vehicle is drawn every \SI{9}{m} of driven path over one \SI{21.7}{s} run, with corridor boundaries, centerline, and driven path overlaid.}
\label{fig:overhead}
\end{figure}

We test the learned certificate in simulation on held-out track geometries and across a range of friction levels, including simultaneous changes in geometry and grip. We compare against certificates re-synthesized for the target geometry or grip condition, quantifying the safety and performance cost of avoiding context-specific synthesis. 
The central finding is that context conditioning makes reachability transferable without making it overly conservative: one certificate maintains containment across the tested contexts while staying close in intervention rate and speed to certificates specialized to the test condition.
We then deploy the resulting certificate as a sampled discrete-time control barrier function filter on a full-scale vehicle at a skidpad (Fig.~\ref{fig:overhead}). The filter is exercised both by a professional driver instructed to approach the boundary and by an autonomous racing controller. Rather than treating absence of violations as the sole measure of success, we characterize both containment and the cost of transfer, including the intervention required to maintain containment and the resulting performance cost.

To the best of our knowledge, this is the first demonstration of context-conditioned HJ reachability for transferable safety certificates across deployment conditions, validated on a full-scale vehicle at the handling limit.

\section{Related Work}

\emph{Vehicle safety at the handling limit.}
At the limit of grip, vehicle safety is commonly enforced by restricting the vehicle to a set of states considered recoverable, either as constraints in predictive control or as triggers for shared control~\cite{beal2013mpc,funke2017collision,brown2017envelopes,erlien2016shared,talbot2024shared}. On full-scale vehicles, Dallas \emph{et al.}~\cite{dallas2025cbf} filter steering and drive commands during drifting using a closed-form safe set with a fixed friction coefficient, identifying adaptation to model uncertainty as an open problem. Closest to our setting, concurrent work learns a control barrier value function by boundary sampling along Pontryagin characteristics and deploys it as a filter for full-scale shared-control drifting~\cite{pmpsampling}. These works demonstrate safety filtering near the handling limit, but their certificates are synthesized for a specific deployment condition.

\emph{Learned reachability and its guarantees.}
Neural reachability methods have made high-dimensional reachable sets tractable~\cite{bansal2021deepreach,singh2025exactbc,feng2025deepreachmpc}, including gradient-free formulations that propagate discrete-time Bellman--Isaacs values without requiring spatial value gradients~\cite{grad_free_reach}. Because a learned value function is an approximation, its zero level set is not a certified safe set; scenario optimization and conformal prediction supply probabilistic correction bounds for learned tubes~\cite{lin2023assurances,lin2024conformal}. Such guarantees are typically formulated for specified dynamics and constraints, so extending them across a continuous context space requires additional treatment. We therefore treat our learned tube as an uncertified certificate, report containment empirically, and regard formal verification as complementary.

\emph{Conditioning and transfer of safety certificates.}
Parameter-conditioned reachability represents families of reachable sets by lifting scalar problem parameters, such as disturbance magnitude or obstacle size, into the state~\cite{borquez2023param,nakamura2023online}. Learned safe sets have also been conditioned on perceptual observations of unknown surroundings, for quadrupedal navigation~\cite{lin2025one} and for simulated racing from ego-vision~\cite{chen2021egovision}. In these works, the conditioning captures aspects of the surrounding environment while the underlying dynamics remain fixed. Our setting couples two distinct forms of conditioning: local boundary geometry determines the safety constraint, while friction and disturbance scale determine the recoverable envelope under the vehicle dynamics. This combination couples changes in the safety constraint with changes in the recoverable envelope within a single learned representation.

On full-scale vehicles, prior reachability-based filtering has used grid-computed sets over low-dimensional state spaces and has been demonstrated in ordinary driving rather than at the handling limit~\cite{leung2020infusing}. Model mismatch has instead typically been addressed by adapting the dynamics, for example through friction estimation~\cite{hsu2010friction,berntorp2020friction} or learned corrections to a nominal model~\cite{kabzan2019lmpc}, or by robustifying the barrier condition against parametric uncertainty~\cite{xu2015robustness,taylor2020adaptive}. These approaches modify the model or margin used by the controller; they do not characterize how a learned reachability certificate transfers when deployment conditions differ from those assumed during synthesis. In contrast, our formulation retains a single learned certificate and exposes the relevant deployment conditions as queryable context, allowing transfer to be evaluated directly against certificates re-synthesized for the target condition. This gap is particularly consequential near the handling limit, where modest changes in grip or vehicle dynamics can change the recoverable set. We therefore evaluate transfer both in simulation, across geometry and grip shifts, and on a full-scale vehicle operating near the friction limit.
\section{Problem Formulation}
\label{sec:problem}

\subsection{Context-indexed safety problems}
\label{sec:problem:system}
We consider a family of safety problems indexed by a context $\context \in \contextset$. The context collects quantities that are fixed over the reachability horizon but vary across deployments. In shared-control racing, these include environment structure, such as track geometry and boundary layout, and operating conditions, such as available grip and disturbance authority.

For a fixed context $c$, the system evolves according to discrete-time dynamics affine in control and disturbance,
\begin{equation}
z_{k+1} = f(z_k, u_k, d_k;\, \context)
        = f_0(z_k;\context) + f_u(z_k;\context) u_k + f_d(z_k;\context)\, d_k,
\label{eq:dyn}
\end{equation}
where $z_k \in \mathcal{Z}$ is the dynamic state, $u_k \in \mathcal{U}$ is the control input, and $d_k \in \mathcal{D}$ is an adversarial disturbance, with all sets compact. 
The adversarial disturbance can represent model uncertainty or external inputs that we want to safeguard against.
The context $\context$ is not controlled by the vehicle and is not treated as an adversarial disturbance; rather, it specifies the particular safety problem being considered.

\subsection{Context-dependent safety specification}
\label{sec:problem:spec}
Safety is specified by a context-dependent margin function $\ell(z;\context)$, with failure set $\mathcal{L}(\context) = \{z \in \mathcal{Z} : \ell(z;\context) \le 0\}$.
The dependence on $\context$ is essential: changing the track geometry changes the lateral-containment constraint and therefore the set of states from which containment is recoverable. 
State bounds beyond those defining $\mathcal{L}(\context)$ are treated as domain limits of the certificate rather than as part of the safety specification.

\subsection{Reachability objective for a fixed context}
\label{sec:problem:brt}
For a fixed context $\context$, the finite-horizon backward reachable tube value function is
\begin{equation}
V^*(z,k;\context) =
\max_{u(\cdot)}
\min_{d(\cdot)}
\left[
\min_{\kappa \in [k,K]}
\ell\left(z^{u,d}_{z,k}(\kappa);\context\right)
\right],
\label{eq:brt}
\end{equation}
where $z^{u,d}_{z,k}(\kappa)$ denotes the trajectory starting from the state $z$ at time $k$ under control sequence $u(\cdot)$, disturbance sequence $d(\cdot)$, and context $\context$. The zero sublevel set of $V^*$ contains the states from which entry into $\mathcal{L}(\context)$ cannot be avoided under worst-case disturbance.

The value function satisfies the Bellman--Isaacs recursion
\begin{equation}
V^*(z,k; \context) = \min\left\{\ell(z; \context),\;
\max_{u \in \mathcal{U}} \min_{d \in \mathcal{D}} V^*\left(f(z,u,d; \context), k{+}1; \context \right)\right\}
\label{eq:recursion}
\end{equation}
with terminal condition $V^*(\cdot,K; \context) = \ell(\cdot; \context)$~\cite{mitchell2005, bansal2017hj}. 

\subsection{Problem statement}
\label{sec:problem:statements}
\begin{problem}[Context-conditioned safety filtering]
\label{prob:filter}

Given a family of context-indexed safety problems $\{f(\cdot;\context),\ell(\cdot;\context)\}_{\context \in \contextset}$, construct a runtime safety filter that, for a queried context $\context_q$ and an arbitrary nominal input $u_{\mathrm{nom}}$, applies an admissible input $u^\star \in \mathcal{U}$ that is minimally modified from $u_{\mathrm{nom}}$ while remaining consistent with the safety certificate for context $\context_q$.
\end{problem}

This problem asks for a safety filter whose behavior adapts to the deployment context. In our setting, this means that the same filtering framework should adapt both to changes in track geometry, which alter the safety constraint, and to changes in grip, which alter the vehicle's recoverable envelope. The next section instantiates the context $\context$ for shared-control racing; Sec.~\ref{sec:approach} then describes how we learn and deploy a single context-conditioned reachable tube for this family of problems.

\section{Vehicle Model and Context}
\label{sec:model}

\subsection{State, control, and context}
\label{sec:model:state}

We model the vehicle in a Frenet frame attached to the centerline of a track indexed by $i$. The track's half-width and curvature profiles are read from lookup tables, denoted $w(i,s)$ and $\kappa(i,s)$. Geometric quantities are referenced to the vehicle center of gravity (CG).

The dynamic state is
\begin{equation}
  z= \begin{bmatrix}
       s & e & \Delta \phi & v & r & \beta & \delta & \tau 
   \end{bmatrix}^\top 
   \label{eq:state}
\end{equation}
where $s$ is arc length, $e$ lateral deviation (positive left of the centerline), $\Delta\phi$ heading error, $v$ velocity, $r$ yaw rate, $\beta$ sideslip, $\delta$ road-wheel steering angle, and $\tau$ commanded longitudinal force. 
The control inputs are rates, $u = [\dot\delta, \dot\tau]^\top$, so actuator bandwidth is included in the safety problem rather than treated as unmodeled lag.

The context for this system is 
$\context = \begin{bmatrix} i & \mu & s_d \end{bmatrix}^\top$
, where $i$ selects the track geometry, $\mu$ is the friction coefficient assumed by the certificate, and $s_d$ scales the authority of the adversarial disturbance. Thus, $i$ changes the track and safety specification, $\mu$ changes the vehicle's available tire force, and $s_d$ selects a robustness budget. Bounds are listed in Table~\ref{tab:params}.

\subsection{Vehicle dynamics} \label{sec:model:vf}
Let $\xi = \Delta\phi + \beta$ be the course angle and abbreviate $\kappa = \kappa(i,s)$. The continuous-time dynamics used inside the discrete-time model are
\begin{alignat}{2}
\dot s &= \frac{v\cos\xi - p\sin\xi}{1 - \kappa e}, &\qquad
\dot v &= \tfrac{1}{m} F_t, \nonumber\\
\dot e &= v\sin\xi + p\cos\xi, &\qquad
\dot r &= \tfrac{1}{I_{zz}} M_z + s_d d_r, \label{eq:vf}\\
\dot{\Delta\phi} &= r - \kappa\, \dot s, &\qquad
\dot\beta &= \tfrac{1}{mv} F_n - r, \nonumber
\end{alignat}
together with $\dot\delta = u_1$, $\dot\tau = u_2$.
Here $F_t$, $F_n$, and $M_z$ are the tangential force, normal force, and yaw moment obtained from the tire model below. 
The disturbance is two-dimensional, $d = (d_r, d_n) \in \mathcal{D} = [-1,1] \times [-0.3,0.3]$, with authority scaled by $s_d$. 

The yaw disturbance captures unmodeled yaw moments, while the normal-velocity disturbance captures uncertainty in lateral velocity or localization.
We write $p = s_d(|v|+1)\,d_n$ for the resulting lateral velocity perturbation. 

\subsection{Tire model and grip context}
\label{sec:model:tires}
Axle forces follow a nonlinear brush tire model~\cite{pacejka2005tire} with longitudinal load transfer and coupling between longitudinal and lateral tire forces. 

The lateral force capacity is 
\begin{equation}
F_y^{\max} = 
\sqrt{(\mu F_z)^2-\zeta F_x^2},
\label{eq:ellipse}
\end{equation}
where $\mu$ is the friction coefficient and $\zeta=0.99$ promotes numerical stability. Lateral forces are then obtained from the brush-tire force--slip relationship~\cite{pacejka2005tire}. Thus, the grip context $\mu$ directly modulates the available tire-force capacity and, consequently, the vehicle dynamics and recoverable safe set.

\subsection{Track context and failure set}
\label{sec:tracks}
Safety is lateral containment between the track boundaries:
\begin{equation}
\ell(z; \context) = w(i,s) - |e|.
\label{eq:ell}
\end{equation}
This margin measures the distance from the vehicle center of gravity to the nearest track boundary, without inflation for the vehicle footprint. 

Because \eqref{eq:ell} is defined by the half-width profile $w(i,s)$, the safety specification changes with the track context. We use a library of \num{38} closed circuits spanning three geometry families: F1Tenth benchmark geometries~\cite{o2020f1tenth}, four full-scale synthesized race-track layouts, and eleven skidpad figures from the physical test site. The library is summarized in Table~\ref{tab:tracks}.

\begin{table}[t]
\caption{Track library. Half-widths are per side; $R_{\min}$ is the minimum radius of curvature. All are single-lap closed loops.}
\label{tab:tracks}
\centering\small
\resizebox{\columnwidth}{!}{
\begin{tabular}{lcccc}
\toprule
Family & Count & Length (m) & Half-width (m) & $R_{\min}$ (m) \\
\midrule
Small-scale circuits & \num{23} & 260--553 & 0.47--2.14 & 0.7--13.7 \\
Full-scale circuits  & \num{4}  & 3190--8188 & 3.55--8.60 & 9.2--21.3 \\
Skidpad figures      & \num{11} & 135--300 & 3.40 (const.) & 9.1--17.0 \\
\bottomrule
\end{tabular}}
\end{table}

\begin{table}[t]
\caption{State, input, and context bounds. 
All values are read from the deployed configuration.
}
\label{tab:params}
\centering\small
\resizebox{0.5\textwidth}{!}{
\begin{tabular}{llr}
\toprule
Symbol & Quantity & Value \\
\midrule
$\delta,\,\dot\delta$ & steering angle, rate & $\pm\num{0.53}$\,rad, $\pm\num{0.6}$\,rad/s \\
$\tau,\,\dot\tau$ & long.\ force, rate & $[-14.1,\,12.7]$\,kN, $\pm\num{12.8}$\,kN/s \\
$v$ & trained speed envelope & $[1.0,\,50.8]$\,m/s \\
$r,\,\beta$ & yaw rate, sideslip & $\pm\num{1.5}$\,rad/s, $\pm\num{0.3}$\,rad \\
$\mu,\,s_d$ & context bounds & $[0.4,\,2.0]$, $[0,\,1]$ \\
$d_r,\,d_n$ & disturbance channels & $\pm\SI{1}{rad/s^2}$, $\pm\num{0.3}$ \\
\bottomrule
\end{tabular}
}
\end{table}

\section{Approach}
\label{sec:approach}

\subsection{Context-conditioned reachable tube}
\label{sec:overall_approach}
Problem~\ref{prob:filter} asks for a safety filter that adapts across a family of context-indexed safety problems. 
A fixed reachable tube would certify one such problem: one track geometry, one assumed friction coefficient, and one disturbance model. We therefore approximate this family jointly with a single context-conditioned value function.
Let $x := (z,\context)$. We define the augmented dynamics:
\begin{equation}
x_{k+1} = \left[f(z_k, u_k, d_k;\, \context); \context \right],
\label{eq:dyn_augmented}
\end{equation}
so the context is held fixed over each reachability rollout. The safety margin is lifted similarly as $\ell(x) = \ell(z;\context)$.

The resulting value function $V_\theta(x,k)$ can be queried at a particular context to obtain the corresponding safety envelope for that geometry, grip, and disturbance assumption. The filter therefore adapts its safety envelope to the queried context rather than enforcing a single envelope that is simultaneously conservative across all contexts.

The context variables play different roles. Friction coefficient and disturbance scale are low-dimensional operating conditions, whereas track geometry is environment structure that must generalize across layouts. We therefore do not provide the track index directly to the value network; instead, we encode the local boundary geometry through an observation map, described next.

\subsection{Observation map and architecture}
\label{sec:obs}
To generalize across tracks, the value network acts on a local observation of the deployment context rather than on the index $i$ directly. The observation is
\begin{equation}
\varphi(x) = \big(\,
\underbrace{B(i,s,e,\Delta\phi,\beta)}_{4 \times 254},\;
\underbrace{e/w(i,s)}_{1},\;
\underbrace{\Delta\phi, v, r, \beta, \delta, \tau}_{6},\;
\underbrace{\mu, s_d}_{2}\,\big),
\label{eq:obs}
\end{equation}
and the learned certificate is $V_\theta(\varphi(x),k)$. We write $V_\theta(x,k)$ for this composition where no ambiguity arises.

Here, $B$ renders the left and right track boundaries as ego-frame polylines, with four channels $(x_L,y_L,x_R,y_R)$ sampled at \SI{1}{m} intervals over a fixed \SI{254}{m} spatial lookahead, corresponding to \SI{5}{s} at the vehicle's maximum speed. The boundaries are reconstructed from the vehicle pose $(i,s,e,\Delta\phi)$ and transformed into the ego frame. A shared one-dimensional convolutional encoder maps $B$ to a 64-dimensional feature vector, which is concatenated with the remaining inputs in $\varphi(x)$ and the curriculum time.

\subsection{Reachability synthesis} 
\label{sec:synthesis}

We approximate the context-conditioned reachable tube using the gradient-free discrete-time solver of~\cite{grad_free_reach}, which propagates the Bellman--Isaacs recursion backward without requiring spatial value gradients. The horizon $T = K\Delta t = \SI{12}{s}$ is discretized at $\Delta t = \SI{0.05}{s}$ and partitioned into $W = 24$ temporal windows of \SI{0.5}{s}. 
Each $\Delta t$ transition is integrated with $n_{\mathrm{sub}} = 5$ explicit Euler substeps. During synthesis, training samples are drawn across the track library and over the conditioning ranges specified in Table~\ref{tab:params} for $\mu$ and $s_d$.

\subsection{Online safety filter} 
\label{sec:filter}

At runtime, the nominal driver or controller proposes an input $u_{\mathrm{nom}}$. The filter queries the learned certificate at the current state and chosen context $c_q$, and applies the closest admissible input that satisfies a discrete-time barrier decrease condition~\cite{agrawal2017dcbf}:
\begin{equation}
V_{\theta}\big(F_{\Delta t_f}(x,u),0\big) \ge
\underbrace{\max\big\{(1 - \gamma \Delta t_f) V_{\theta}(x,0),\; 0\big\}}
_{\textstyle V_{\mathrm{req}}(x;\gamma)},
\label{eq:dcbf}
\end{equation}
with $\gamma = \num{3.0}$. Here, $F_{\Delta t_f}$ denotes the state predicted $\Delta t_f$ seconds ahead under the control input $u$. Because $V_\theta$ is never differentiated, \eqref{eq:dcbf} is evaluated over a sampled candidate set $\mathcal{U}_{\mathrm{batch}}$ in one batched forward pass, giving $\mathcal{U}_{\mathrm{safe}}(x)$ -- the set of control inputs that satisfy the CBF condition. 
The batch contains
$|\mathcal{U}_{\mathrm{batch}}| = 129$ candidates comprising the nominal action, control-box vertices, Gaussian and uniform perturbations, and the co-trained recovery action. The applied control is
\begin{equation}
u^\star =
\begin{cases}
\displaystyle\argmin_{u \in \mathcal{U}_{\mathrm{safe}}(x)}
  \big\|\Lambda (u - u_{\mathrm{nom}})\big\|_2,
  & \mathcal{U}_{\mathrm{safe}}(x) \neq \emptyset,\\[4pt]
\pi^u_\phi(x), & \text{otherwise},
\end{cases}
\label{eq:filter}
\end{equation}
where $\Lambda = \mathrm{diag}(1/\dot\delta_{\max}, 1/\dot\tau_{\max})$ normalizes each channel by its available authority. The fallback $\pi^u_\phi$ is the recovery policy co-trained during value function synthesis; it shares the observation encoder with $V_\theta$ and emits bang-bang actions. 
\section{Simulation Study} 
\label{sec:sim}

Our simulation study evaluates whether one context-conditioned certificate can replace separate certificates synthesized for each track and grip condition. We ask four questions: (i) whether the same certificate generalizes to unseen track geometries, (ii) whether it transfers across the conditioned grip range, (iii) whether geometry and grip shifts compound when they occur together, and (iv) what happens when the certificate is queried at the wrong grip.

\subsection{Simulation Setup}
\label{sec:sim:protocol}

We simulate the vehicle model of Sec.~\ref{sec:model} with a plant friction coefficient $\mu_{\mathrm{true}}$. The safety filter is queried at $\mu_q$, which equals $\mu_{\mathrm{true}}$ unless stated otherwise. The nominal controller is a curvature-feedforward racing-line tracker with PD lateral and PI speed feedback. It is deliberately detuned so that, without filtering, it overdrives the tires and leaves the corridor. The same nominal controller, initial conditions, and filter parameters are used in every comparison.

\noindent \textit{\textbf{Compared configurations.}}
\textsc{NoFilter} is the nominal controller alone. \textsc{Ours} is a single certificate conditioned on track geometry and grip and queried online. We also report context-specialized references: \textsc{NominalGrip}, synthesized at the nominal grip; \textsc{OneTrack}, synthesized separately for each test geometry; and \textsc{OneGrip}, synthesized separately for each test grip. These references quantify the performance cost of avoiding context-specific re-synthesis. Finally, \textsc{RobustGrip} evaluates the certificate as $\min_{\mu \in [\mu_{\mathrm{lo}}, \mu_{\mathrm{hi}}]} V_\theta(x;\mu)$, taking the pointwise minimum over a $\pm10\%$ band of assumed grip.

\noindent \textit{\textbf{Metrics.}}
We report three quantities. \emph{Violations} is the percentage of trials in which $\ell(x_t) < 0$ at any time; a rollout terminates on violation. \emph{Intervention} is the fraction of control steps at which the filtered input differs from the nominal input, measuring the cost imposed by the filter. \emph{Speed} is the mean longitudinal speed over completed trials only, and is reported as \textit{n/a} when no trials complete.

\noindent \textit{\textbf{Evaluation details.}}
For in-distribution evaluation, the network is trained on the \num{11} evaluation skidpad tracks; for out-of-distribution evaluation, it is trained on the \num{27} racing tracks and evaluated on the \num{11} skidpad geometries. The training bounds for $\mu$ and $s_d$ are given in Table~\ref{tab:params}. Unless otherwise stated, each simulation operating condition is evaluated over \num{30} independent rollouts.

\subsection{The cost of querying the wrong grip}
\label{sec:sim:mismatch}
\begin{figure}[t]
\centering
\includegraphics[width=\columnwidth]{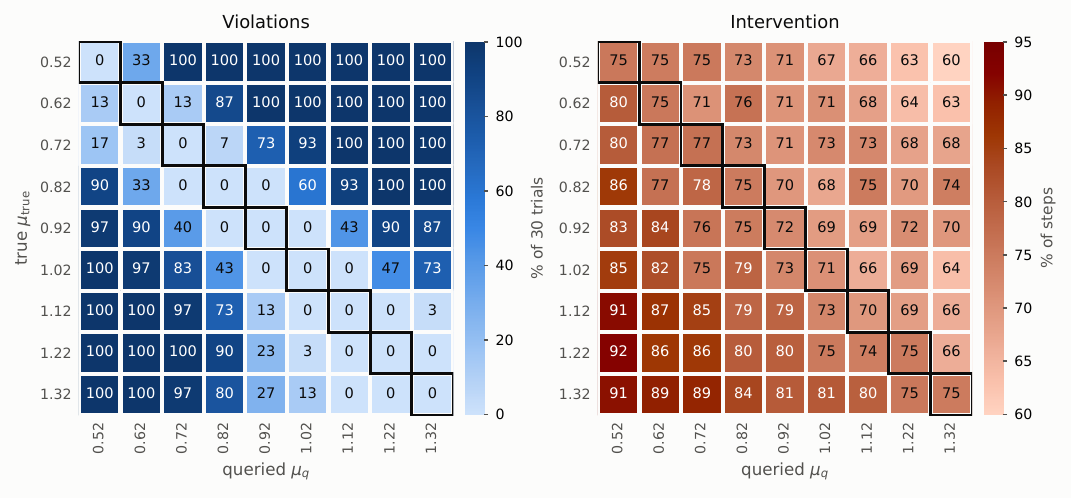}
\vspace{-2em}
\caption{Assumed-grip mismatch. Rows are the plant grip $\mu_{\mathrm{true}}$, columns are the queried grip $\mu_q$, and boxed cells are matched. \emph{Left:} violations over \num{30} trials. \emph{Right:} intervention rate.}
\vspace{-1em}
\label{fig:mismatch}
\end{figure}

We sweep the plant grip $\mu_{\mathrm{true}}$ and queried grip $\mu_q$ independently over $[0.52,1.32]$, evaluating all \num{81} combinations with \num{30} trials each (Fig.~\ref{fig:mismatch}).
At every grip from $0.52$ to $1.32$, a correctly queried certificate completes all \num{30} trials. Off the diagonal, safety containment remains intact for mismatches of roughly $\pm 0.10$ in $\mu$, after which violations increase rapidly. At the identified grip $\mu_0 = 0.92$, querying $\mu$ of $0.20$ too low or too high raises violations to \SI{40}{\percent} and \SI{43}{\percent}, respectively; at $|\mu_q-\mu_{\mathrm{true}}|=0.30$, violations reach roughly \SI{90}{\percent} in both directions.

The two sides fail through different mechanisms. Intervention falls monotonically as the queried grip rises, from \SI{92}{\percent} at the most pessimistic query to \SI{60}{\percent} at the most optimistic. An optimistic certificate credits the vehicle with margin it does not have, so the filter intervenes less and the vehicle leaves the corridor. A pessimistic certificate does the opposite and still fails at large mismatch: it intervenes as though grip were scarce, and the resulting conservative commands can also produce loss of containment when the vehicle has more grip than assumed.

The relative severity depends on the true grip. Near $\mu_0$, the two mismatch directions produce similar violation rates. At low grip, however, overestimating friction is substantially more dangerous: at $\mu_{\mathrm{true}} = 0.72$, querying $0.92$ gives \SI{73}{\percent} violations, compared with \SI{17}{\percent} when querying $0.52$. Thus, the most consequential failure mode in this sweep is optimistic grip estimation under already-limited adhesion.
Overall, these results suggest the importance of context adaptation to ensure vehicle safety.

\subsection{Generalization across geometry}
\label{sec:sim:track}
We next isolate the geometry axis.
We hold grip at the identified $\mu_0 = 0.92$ and evaluate on the same 11 geometries used to define the in-distribution and out-of-distribution training splits. For the out-of-distribution model, these geometries are absent from the training library.
Table~\ref{tab:trackgen} reports the results.

\begin{table}[t]
\centering\small
\caption{Generalization across held-out geometry. 11 held-out tracks, 30 trials each, $\mu_q = \mu_{\mathrm{true}} = 0.92$, $s_d = 0.5$. \textsc{NoFilter} completes no trial, so its speed is not a progress measure and is omitted. \textsc{OneTrack} is a geometry-specialized upper bound re-synthesized for each test track.}
\label{tab:trackgen}
\centering
\begin{tabular}{lccc}
\toprule
Configuration & Viol.\ (\%) & Interv.\ (\%) & Speed (m/s) \\
\midrule
\textsc{NoFilter}    & 100.0   & n/a      & n/a \\
\textsc{RobustGrip}  & 0.0 & 72.3 & 12.66 \\
\textsc{Ours}        & 0.0     & 74.2     & 13.07 \\
\midrule
\textsc{OneTrack}    & 0.0 &  74.1 &  13.16 \\
\bottomrule
\end{tabular}
\vspace{-1em}
\end{table}

The nominal controller leaves the corridor on every trial. 
The context-conditioned filter removes all violations on geometries it was never trained on. Its cost is also nearly identical to the geometry-specialized upper bound: \textsc{Ours} intervenes only 0.1 percentage points more often than \textsc{OneTrack} and runs within 0.09$\mathrm{m/s}$. Thus, in this setting, conditioning on local boundary geometry allows one certificate to generalize across tracks with almost no loss relative to re-synthesizing on the test geometry.

\subsection{Generalization across grip}
\label{sec:sim:grip}

We now isolate the grip axis. The geometries are drawn from the training library, so the only shift is in grip. We sweep the plant grip across the conditioning range while querying the certificate at $\mu_q=\mu_{\mathrm{true}}$.

\begin{table}[t]
\centering\small
\caption{Generalization across grip, in-distribution geometry. Columns are plant grip, with $\mu_q=\mu_{\mathrm{true}}$ and $s_d=0.5$. 30 trials per cell on 11 training geometries. Only \textsc{NoFilter} and \textsc{NominalGrip} incur violations; all other configurations have zero violations. \textsc{OneGrip} is a grip-specialized upper bound re-synthesized at each column's grip value.}
\label{tab:gripgen}
\centering
\resizebox{\columnwidth}{!}{
\begin{tabular}{lcccccc}
\toprule
$\mu_{\mathrm{true}}$ & $0.4$ & $0.6$ & $0.92$ & $1.2$ & $1.6$ & $2.0$ \\
\midrule
\multicolumn{7}{l}{\emph{Violations} (\%)} \\
\textsc{NoFilter}    & 100.0 & 100.0 & 100.0 & 100.0 & 100.0 & 100.0 \\
\textsc{NominalGrip} & 100.0 & 100.0 & 0.0 & 33.3 & 56.7 & 53.3 \\
\hline
\hline
\multicolumn{7}{l}{\emph{Intervention rate} (\%)} \\
\textsc{NominalGrip} & n/a & n/a & 72.4 & 76.3 & 78.8 & 88.0 \\
\textsc{RobustGrip}  & 76.3 & 74.3 & 71.6 & 72.1 & 74.3 & 75.0 \\
\textsc{Ours}        & 77.5 & 73.2 & 73.6 & 69.9 & 69.9 & 68.9 \\
\midrule
\textsc{OneGrip}     & 75.3 & 72.5 & 72.4 & 68.7 & 68.5 & 66.9 \\
\hline
\hline
\multicolumn{7}{l}{\emph{Speed} (m/s)} \\
\textsc{NominalGrip} & n/a & n/a & 13.41 & 14.16 & 14.88 & 17.22 \\
\textsc{RobustGrip}  & 5.78 & 8.72 & 13.17 & 15.02 & 17.12 & 18.84 \\
\textsc{Ours}        & 5.77 & 9.53 & 13.27 & 15.06 & 17.04 & 19.01 \\
\midrule
\textsc{OneGrip}     & 5.86 & 9.72 & 13.41 & 15.17 & 17.23 & 19.17 \\
\bottomrule
\end{tabular}}
\end{table}

The results first show the importance of grip conditioning. \textsc{NominalGrip}, synthesized only at $\mu=0.92$, maintains containment at the nominal grip but incurs substantial violations away from it. In contrast, \textsc{Ours} maintains zero violations across the full conditioned range, matching the specialist \textsc{OneGrip} at every operating point.

The specialist recovers only a small amount of additional performance. \textsc{Ours} intervenes \SIrange{0.7}{2.2}{} percentage points more often and runs \SIrange{0.7}{2}{\percent} slower than \textsc{OneGrip} across the sweep. This quantifies the performance cost of representing the full $\mu\in[0.4,2.0]$ range with one value function rather than re-synthesizing a certificate at each operating point. \textsc{RobustGrip} also maintains zero violations, but its pointwise minimum over a $\pm10\%$ grip band incurs a larger intervention and speed cost than \textsc{Ours} at several operating points.

\subsection{Simultaneous geometry and grip shift}
\label{sec:sim:cross}
Finally, we shift geometry and grip together by running the grip sweep of Table~\ref{tab:gripgen} on the 11 held-out geometries. This is the most direct test of the context-conditioned claim: the certificate must adapt to an unseen track while also being queried at different grip levels.

\begin{table}[t]
\centering\small
\caption{Simultaneous geometry and grip shift. $\mu_q=\mu_{\mathrm{true}}$, $s_d=0.5$, and 30 trials are used per cell. Only \textsc{NoFilter} and \textsc{NominalGrip} incur violations; all other configurations have zero violations.}
\label{tab:cross}
\resizebox{\columnwidth}{!}{
\begin{tabular}{lcccccc}
\toprule
$\mu_{\mathrm{true}}$ & $0.4$ & $0.6$ & $0.92$ & $1.2$ & $1.6$ & $2.0$ \\
\midrule
\multicolumn{7}{l}{\emph{Violations} (\%)} \\
\textsc{NoFilter}    & 100.0 & 100.0 & 100.0 & 100.0 & 100.0 & 100.0 \\
\textsc{NominalGrip} & 100.0 & 100.0 & 0.0 & 43.3 & 53.3 & 53.3 \\
\midrule
\multicolumn{7}{l}{\emph{Intervention rate} (\%)} \\
\textsc{NominalGrip} & n/a & n/a & 72.8 & 77.8 & 80.7 & 90.4 \\

\textsc{OneTrack}    & 77.9 & 75.1 & 74.1 & 71.7 & 70.5 & 68.8 \\
\textsc{OneGrip}     & 77.1 & 73.2 & 72.8 & 69.2 & 69.1 & 67.3 \\
\textsc{RobustGrip}  & 78.1 & 76.3 & 72.3 & 72.8 & 73.7 & 75.9 \\
\textsc{Ours}        & 78.3 & 74.1 & 74.2 & 70.5 & 70.4 & 69.9  \\
\midrule
\multicolumn{7}{l}{\emph{Speed} (m/s)} \\
\textsc{NominalGrip} & n/a & n/a & 13.39 & 13.62 & 13.89 & 16.91 \\
\textsc{OneTrack}    & 5.30 & 8.99 & 13.16 & 14.99 & 16.91 & 18.98 \\
\textsc{OneGrip}     & 5.66 & 9.32 & 13.39 & 15.14 & 17.29 & 19.13 \\
\textsc{RobustGrip}  & 5.15 & 8.21 & 12.66 & 14.87 & 16.75 & 18.61 \\
\textsc{Ours}        & 5.32 & 8.93 & 13.07 & 14.96 & 16.84 & 18.91 \\
\bottomrule
\end{tabular}}
\vspace{-1em}
\end{table}

Table~\ref{tab:cross} shows that the two shifts do not compound into failures: \textsc{Ours} maintains containment for every held-out geometry and every grip value. Its cost remains close to the geometry-specialized reference, staying within \SI{1.2} percentage points of intervention and \SI{0.09}{m/s} of \textsc{OneTrack} at every grip. The comparison to \textsc{OneGrip} is more revealing: \textsc{Ours} trails by up to \SI{2.6} intervention points and \SI{0.45}{m/s} in speed, suggesting that grip is the harder axis to cover than geometry. This is consistent with the model: $\mu$ directly scales the tire-force capacity and the size of the recoverable set, whereas geometry enters through the boundary observation, over which the network generalizes well.

Overall, the simulation results show that context conditioning provides the desired tradeoff. A single certificate adapts across unseen geometry and a wide range of grip values, with zero violations in the tested matched-context conditions and only modest cost relative to certificates specialized to the test context.

\section{Hardware Deployment}
\label{sec:results}

\subsection{Experimental Setup}
\label{sec:setup}

\subsubsection{Vehicle, site, and surface}

Experiments were conducted on a closed skidpad on a full-scale instrumented electric passenger vehicle. A dual-antenna RTK GPS/INS system supplies state estimates at \SI{200}{Hz}, and the filter runs on an onboard NVIDIA GeForce RTX 4070 Ti SUPER GPU with an Intel Core i9-13900E (13th Gen) processor. Every session ran on dry pavement, with no precipitation in the preceding \SI{24}{h}. Modeled surface temperature spanned \SIrange{27.6}{48.3}{\celsius} across the campaign.

\subsubsection{Track library and conditions}

We synthesize three value functions, which differ only in their training geometries. \textsc{InD} is trained on the \num{11} skidpad geometries, which include the deployment geometry. \textsc{OOD} is trained on \num{27} racing circuits, which exclude it, so the deployed centerline is held out. \textsc{Finetuned} starts from \textsc{OOD} and is then adapted on data collected at the deployment geometry. The model, horizon, and filter are identical throughout.

\begin{figure*}[t]
\centering
\includegraphics[width=0.75\textwidth]{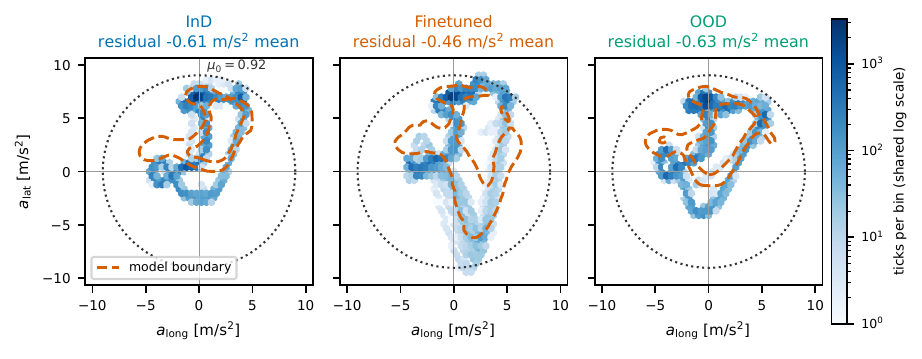}
\vspace{-1em}
\caption{Measured and model-implied acceleration envelopes for the three autonomous tracker sessions. Both channels are smoothed with a \SI{0.1}{s} moving average. Solid curves show the measured envelope and dashed curves the envelope predicted by the vehicle model used by the certificate; the $\mu_0=0.92$ friction circle is shown for reference.}
\vspace{-1em}
\label{fig:ctrlgg}
\end{figure*}

\subsection{Adversarial Driving with Safety Filtering}
\label{sec:adv_exp}

A professional driver was instructed to drive the vehicle out of the corridor, providing adversarial commands that the filter could oppose. All reported results are on the oval skidpad at the identified grip $\mu_0=0.92$. 

\subsubsection{Operating regime}

Accelerations come from the vehicle's IMU sensors, smoothed with a \SI{0.1}{s} moving average. The 99th-percentile acceleration magnitude across all sessions is \SI{0.99}{g}, and the top speed is \SI{20.8}{m/s}. The vehicle reached this acceleration regime without developing large sideslip. Sideslip at the \num{99.9}th percentile stays below \ang{6.6} in the sessions with clean $\beta$ estimation.

\subsubsection{Where the filter engages}

\begin{figure}[t]
\centering
\vspace{-1em}
\includegraphics[width=\columnwidth]{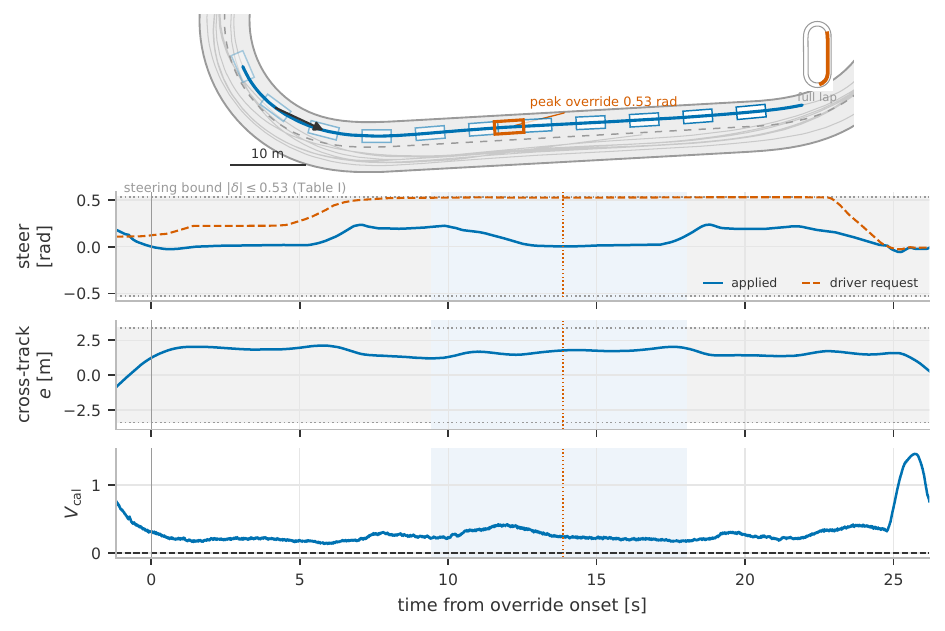}
\vspace{-1em}
\caption{One sustained override event. \emph{Top:} where it happened, with the vehicle drawn through the corner and the instant of peak override boxed; inset shows the position on the full lap. \emph{Below:} the driver's steering request against the angle actually applied, cross-track error against the corridor edges.}
\vspace{-1em}
\label{fig:filter_path}
\end{figure}

Fig.~\ref{fig:filter_path} shows the filter at work during one sustained override. Over moving samples, it modifies the driver's command \SI{84.1}{\percent} of the time. Intervention runs at \SI{99}{\percent} near $V_{\theta}=0$ and falls to \SI{15}{\percent} in the highest decile of $V_{\theta}$. On intervened samples at low $V_{\theta}$, the applied steering shares the driver's sign only \SIrange{12}{37}{\percent} of the time. Thus, near the boundary, the filter often reverses rather than merely attenuates the driver's steering command.

Intervention remains between \SI{81}{\percent} and \SI{94}{\percent} across all curvature bins, from straight to the tightest corner. Curvature instead affects correction magnitude, which grows by roughly a factor of two on the torque channel between the two extremes.

\subsubsection{Behavior across queried grip}
\label{sec:adv:grip}

We next test queried-grip mismatch on hardware, targeting the zero-violation band identified in Fig.~\ref{fig:mismatch}. We repeated the protocol with the certificate queried at $\mu_q \in {0.80, 0.85, 0.92, 1.00}$ on a surface delivering the identified $\mu_0 = 0.92$.

Containment held at every queried value, including runs that reached accelerations beyond the envelope implied by $\mu_0=0.92$. The hardware therefore confirms containment within the interior of the simulated robustness band rather than at its edges: the first simulated failures occur at the next tested mismatch levels, corresponding to querying $\mu_q=0.72$ and $\mu_q=1.12$, which were not evaluated on hardware.

Intervention changes in the direction predicted by simulation, but run-to-run variability dominates the effect: two consecutive drives at $\mu_q=0.85$ differ by \SI{23}{} percentage points, compared with a \SI{4}{}-point change between $\mu_q=0.92$ and $0.80$. We therefore use the hardware experiment to establish containment under queried-grip mismatch rather than to characterize its intervention-cost trend.

\subsection{Safety Filtering on an Autonomous Racing Controller}
\label{sec:ctrl_exp}

The adversarial protocol exercises the certificate but does not quantify its performance cost, since it does not involve timed laps. We therefore also evaluate the filter with an autonomous racing controller: the nominal tracking controller of Sec.~\ref{sec:sim:protocol}, following a minimum-curvature racing line. This experiment extends the evaluation in Table~\ref{tab:trackgen} to hardware, allowing us to quantify the performance cost of deploying the certificate on a geometry absent from its training set.

\subsubsection{Two geometries, three certificates}

We drove two skidpad figures. The oval is a loop of constant-radius turns and short straights; the bean is an asymmetric closed curve whose curvature varies continuously. Table~\ref{tab:trio} reports every distinct drive on the two.

\begin{table}[t]
\caption{Autonomous racing controller on two geometries. $\mu_q = 0.92$ and $s_d = 0.6$; only the value function differs. Acceleration is the 99th percentile of the acceleration-magnitude samples after a \SI{0.1}{s} moving average, computed over 10 laps per configuration.}
\label{tab:trio}
\centering
\resizebox{\columnwidth}{!}{
\begin{tabular}{llccccc}
\toprule
Geometry & Model & Lap (s) & $\bar v$ (m/s) & $v_{\max}$ (m/s) & $p_{99}$ $g$ \\
\midrule
Oval & \textsc{InD} & 18.43 $\pm$ 0.07 & 12.80 & 17.14 & 0.870 \\
Oval & \textsc{Finetuned} & 17.85 $\pm$ 0.07 & 13.15 & 18.12 & 0.939 \\
Oval & \textsc{OOD} & 18.08 $\pm$ 0.04 & 12.94 & 18.43 & 0.803 \\
\midrule
Bean & \textsc{InD} & 22.38 $\pm$ 0.13 & 13.57 & 19.27 & 0.965 \\
Bean & \textsc{OOD} & 22.30 $\pm$ 0.09 & 13.66 & 19.03 & 0.890 \\
\bottomrule
\end{tabular}}
\vspace{-1em}
\end{table}

On the bean, the held-out and in-distribution certificates produce nearly identical performance. Mean lap time differs by \SI{0.08}{s}, mean speed by \SI{0.09}{m/s}, and intervention by \SI{0.1} percentage points. On the oval, \textsc{OOD} has a mean lap time \SI{0.35}{s} shorter than \textsc{InD} and \SI{0.23}{s} longer than \textsc{Finetuned}; the three certificates span \SI{3.1}{\percent} in mean lap time, with \textsc{OOD} also reaching the highest peak speed at \SI{18.43}{m/s}. Consistent with the simulation results in Table~\ref{tab:trackgen}, the held-out certificate therefore incurs little additional performance cost relative to certificates specialized to the deployment geometry.

Figure~\ref{fig:ctrlgg} also exposes a model mismatch that is not visible from containment alone. The measured acceleration envelope does not coincide with the envelope predicted by the model used by the certificate. The model overpredicts the measured envelope over much of the operating range but underpredicts it near the handling limit. The discrepancy is strongly directional: the largest gap occurs near \SI{-90}{\degree}, corresponding to the loaded lateral direction on the counterclockwise oval. The maximum directional gap is \SI{3.22}{m/s^2} for \textsc{Finetuned}, \SI{2.91}{m/s^2} for \textsc{OOD}, and \SI{1.92}{m/s^2} for \textsc{InD}. Thus, the deployment demonstrates containment despite model mismatch in the high-load regime relevant to the experiment.
\section{Conclusion}

We presented a context-conditioned neural Hamilton--Jacobi reachable tube that represents a family of safety problems across track geometry, friction, and disturbance conditions, and deployed it as a runtime safety filter on a full-scale vehicle near the friction limit. In simulation, a single value function transferred across held-out geometries and the conditioned friction range without re-synthesis, including simultaneous geometry and grip shifts. On hardware, lateral containment held in every session, including trials in which a professional driver deliberately attempted to leave the corridor.

The hardware results further demonstrate that this transfer can be achieved with modest performance cost. Over ten autonomous laps under a fixed racing-line tracker, lap time varied by \SI{3.1}{\percent} across certificates trained on in-distribution, held-out, and fine-tuned geometry sets. The held-out and in-distribution certificates also produced comparable performance on the two hardware geometries, both of which were absent from the held-out certificate's training library. Together with the simulation results, these experiments show that a single context-conditioned certificate can transfer across deployment geometries and operating conditions without the need for context-specific re-synthesis.

%
Several limitations remain. The certified property is lateral containment of the vehicle center of gravity; rollover, thermal degradation, tire wear, and containment of the full vehicle footprint are outside the specification. The full-scale evaluation is also limited to two skidpad geometries---an oval and a bean-shaped course. Thus, while the simulation study evaluates transfer across a broader geometry library, the hardware results should be read as deployment validation rather than as evidence of transfer across all track families. Finally, the hardware mismatch experiments test only an interior range of queried grip values: containment held over the tested range, but the first failures observed in simulation occurred at the next mismatch levels, which were not evaluated on the vehicle. Extending the hardware study to lower-grip surfaces and diverse full-scale geometries would provide a more complete characterization of context-conditioned certificate transfer.

\section{Acknowledgments}
This research is supported by the TRI University Research Program (URP) 3.0.

\bibliographystyle{IEEEtran}
\bibliography{references}

@article{mitchell2005,
  author={Mitchell, Ian M. and Bayen, Alexandre M. and Tomlin, Claire J.},
  title={A Time-Dependent Hamilton-Jacobi Formulation of Reachable Sets for Continuous Dynamic Games},
  journal={IEEE Transactions on Automatic Control},
  volume={50}, number={7}, pages={947--957}, year={2005}
}

@inproceedings{bansal2017hj,
  author={Bansal, Somil and Chen, Mo and Herbert, Sylvia and Tomlin, Claire J.},
  title={Hamilton-Jacobi Reachability: A Brief Overview and Recent Advances},
  booktitle={IEEE Conference on Decision and Control (CDC)},
  pages={2242--2253}, year={2017}
}

@inproceedings{bansal2021deepreach,
  author={Bansal, Somil and Tomlin, Claire J.},
  title={{DeepReach}: A Deep Learning Approach to High-Dimensional Reachability},
  booktitle={IEEE International Conference on Robotics and Automation (ICRA)},
  pages={1817--1824}, year={2021}
}

@inproceedings{lin2023assurances,
  author={Lin, Albert and Bansal, Somil},
  title={Generating Formal Safety Assurances for High-Dimensional Reachability},
  booktitle={IEEE International Conference on Robotics and Automation (ICRA)},
  pages={10525--10531}, year={2023}
}

@inproceedings{lin2024conformal,
  author={Lin, Albert and Bansal, Somil},
  title={Verification of Neural Reachable Tubes via Scenario Optimization and Conformal Prediction},
  booktitle={Learning for Dynamics and Control (L4DC)},
  series={PMLR}, volume={242}, pages={719--731}, year={2024}
}

@inproceedings{singh2025exactbc,
  author={Singh, Aditya and Feng, Zeyuan and Bansal, Somil},
  title={Exact Imposition of Safety Boundary Conditions in Neural Reachable Tubes},
  booktitle={IEEE International Conference on Robotics and Automation (ICRA)},
  year={2025}
}

@inproceedings{borquez2023param,
  author={Borquez, Javier and Nakamura, Kensuke and Bansal, Somil},
  title={Parameter-Conditioned Reachable Sets for Updating Safety Assurances Online},
  booktitle={IEEE International Conference on Robotics and Automation (ICRA)},
  year={2023}
}

@inproceedings{nakamura2023online,
  author={Nakamura, Kensuke and Bansal, Somil},
  title={Online Update of Safety Assurances Using Confidence-Based Predictions},
  booktitle={IEEE International Conference on Robotics and Automation (ICRA)},
  year={2023}
}

@article{leung2020infusing,
  author={Leung, Karen and Schmerling, Edward and Zhang, Mengxuan and Chen, Mo and Talbot, John and Gerdes, J. Christian and Pavone, Marco},
  title={On Infusing Reachability-Based Safety Assurance within Planning Frameworks for Human-Robot Vehicle Interactions},
  journal={The International Journal of Robotics Research},
  volume={39}, number={10-11}, pages={1326--1345}, year={2020}
}

@misc{chen2021egovision,
  author={Chen, Bingqing and Francis, Jonathan and Oh, Jean and Nyberg, Eric and Herbert, Sylvia L.},
  title={Safe Autonomous Racing via Approximate Reachability on Ego-Vision},
  note={arXiv:2110.07699}, year={2021}
}

@article{beal2013mpc,
  title={Model predictive control for vehicle stabilization at the limits of handling},
  author={Beal, Craig Earl and Gerdes, J Christian},
  journal={IEEE Transactions on Control Systems Technology},
  volume={21},
  number={4},
  pages={1258--1269},
  year={2012},
  publisher={IEEE}
}

@article{erlien2016shared,
  author={Erlien, Stephen M. and Fujita, Susumu and Gerdes, J. Christian},
  title={Shared Steering Control Using Safe Envelopes for Obstacle Avoidance and Vehicle Stability},
  journal={IEEE Transactions on Intelligent Transportation Systems},
  volume={17}, number={2}, pages={441--451}, year={2016}
}

@article{funke2017collision,
  author={Funke, Joseph and Brown, Matthew and Erlien, Stephen M. and Gerdes, J. Christian},
  title={Collision Avoidance and Stabilization for Autonomous Vehicles in Emergency Scenarios},
  journal={IEEE Transactions on Control Systems Technology},
  volume={25}, number={4}, pages={1204--1216}, year={2017}
}

@article{brown2017envelopes,
  author={Brown, Matthew and Funke, Joseph and Erlien, Stephen and Gerdes, J. Christian},
  title={Safe Driving Envelopes for Path Tracking in Autonomous Vehicles},
  journal={Control Engineering Practice},
  volume={61}, pages={307--316}, year={2017}
}

@article{talbot2024shared,
  author={Talbot, John and Brown, Matthew and Gerdes, J. Christian},
  title={Shared Control Up to the Limits of Vehicle Handling},
  journal={IEEE Transactions on Intelligent Vehicles},
  volume={9}, number={1}, pages={2977--2987}, year={2024}
}

@inproceedings{dallas2025cbf,
  author={Dallas, James and Talbot, John and Suminaka, Michael and Thompson, Michael and Lew, Thomas and Orosz, G{\'a}bor and Subosits, John},
  title={Control Barrier Functions for Shared Control and Vehicle Safety},
  booktitle={American Control Conference (ACC)},
  pages={4203--4210}, year={2025}
}

@inproceedings{agrawal2017dcbf,
  author={Agrawal, Ayush and Sreenath, Koushil},
  title={Discrete Control Barrier Functions for Safety-Critical Control of Discrete Systems with Application to Bipedal Robot Navigation},
  booktitle={Robotics: Science and Systems (RSS)}, year={2017}
}

@inproceedings{taylor2020adaptive,
  author={Taylor, Andrew J. and Ames, Aaron D.},
  title={Adaptive Safety with Control Barrier Functions},
  booktitle={American Control Conference (ACC)}, pages={1399--1405}, year={2020}
}

@inproceedings{xu2015robustness,
  author={Xu, Xiangru and Tabuada, Paulo and Grizzle, Jessy W. and Ames, Aaron D.},
  title={Robustness of Control Barrier Functions for Safety Critical Control},
  booktitle={IFAC Conference on Analysis and Design of Hybrid Systems (ADHS)},
  volume={48}, number={27}, pages={54--61}, year={2015}
}

@article{hsu2010friction,
  author={Hsu, Yung-Hsiang Judy and Laws, Shad M. and Gerdes, J. Christian},
  title={Estimation of Tire Slip Angle and Friction Limits Using Steering Torque},
  journal={IEEE Transactions on Control Systems Technology},
  volume={18}, number={4}, pages={896--907}, year={2010}
}

@article{berntorp2020friction,
  title={Online Bayesian tire-friction learning by Gaussian-process state-space models},
  author={Berntorp, Karl},
  journal={IFAC-PapersOnLine},
  volume={53},
  number={2},
  pages={13939--13944},
  year={2020},
  publisher={Elsevier}
}

@article{kabzan2019lmpc,
  author={Kabzan, Juraj and Hewing, Lukas and Liniger, Alexander and Zeilinger, Melanie N.},
  title={Learning-Based Model Predictive Control for Autonomous Racing},
  journal={IEEE Robotics and Automation Letters},
  volume={4}, number={4}, pages={3363--3370}, year={2019}
}

@inproceedings{grad_free_reach,
  title     = {Gradient-Free Neural Hamilton-Jacobi Reachability for Scalable Safety-Critical Control},
  author    = {Feng, Z. and Sahin, A. F. and Thorup, S. and Bansal, S.},
  booktitle = {Conference on Robot Learning (CoRL)},
  year      = {2026}
}

@article{pmpsampling,
  title={Boundary Sampling to Learn Predictive Safety Filters via Pontryagin's Maximum Principle},
  author={Dallas, James and Lew, Thomas and Talbot, John and DeCastro, Jonathan and Bansal, Somil and Subosits, John},
  journal={arXiv preprint arXiv:2604.13325},
  year={2026}
}

@inproceedings{feng2025deepreachmpc,
  author={Feng, Zeyuan and Qiu, Le and Bansal, Somil},
  title={Bridging Model Predictive Control and Deep Learning for Scalable Reachability Analysis},
  booktitle={Robotics: Science and Systems (RSS)}, year={2025}
}

@article{o2020f1tenth,
  title={F1tenth: An open-source evaluation environment for continuous control and reinforcement learning},
  author={O'Kelly, Matthew and Zheng, Hongrui and Karthik, Dhruv and Mangharam, Rahul},
  journal={Proceedings of Machine Learning Research},
  volume={123},
  year={2020}
}

@article{lin2025one,
  title={One filter to deploy them all: Robust safety for quadrupedal navigation in unknown environments},
  author={Lin, Albert and Peng, Shuang and Bansal, Somil},
  journal={IEEE Transactions on Robotics},
  volume={42},
  pages={545--560},
  year={2025},
  publisher={IEEE}
}

@book{pacejka2005tire,
  title={Tire and vehicle dynamics},
  author={Pacejka, Hans},
  year={2005},
  publisher={Elsevier}
}

\end{document}